\documentclass[a4paper,11pt]{article}
\usepackage{amsmath,amssymb}
\usepackage{graphicx}
\usepackage{hyperref}
\usepackage[left=1in,right=1in,top=1in,bottom=1in]{geometry}

\newcommand{\be}{\begin{equation}} \newcommand{\ee}{\end{equation}}
\newcommand{\ba}{\begin{eqnarray}} \newcommand{\ea}{\end{eqnarray}}

\title{Parity in Planck full-mission CMB temperature maps}
\author{Srikanta Panda$^1$, Pavan K. Aluri$^2$, Pramoda Kumar Samal$^{1,3}$, Pranati K. Rath$^4$}
\date{}

\begin{document}
\maketitle
\begin{center}
$^1$ Department of Physics, Utkal University, Bhubaneswar-751004, Odisha, India\\
$^2$ Department of Physics, IIT(BHU), Varanasi-221005, Uttar Pradesh, India\\
$^3$ School of Physics, Gangadhar Meher University, Sambalpur-768004, Odisha, India\\
$^4$ Department of Physics, Khallikote (Autonomous) College, Berhampur University, Berhampur-760001, Odisha, India
\end{center}

\begin{abstract}

In the standard model of cosmology, Cosmic Microwave Background (CMB) sky is expected
to show no symmetry preferences. Following our previous studies, we explore the presence 
of any particular parity preference in the latest full-mission CMB temperature maps
from ESA's Planck probe. Specifically, in this work, we will probe (a)symmetry in power 
between even and odd multipoles of CMB via it's angular power spectrum from Planck 2015
data. Further we also assess any specific preference for mirror parity (a)symmetry,
by analysing the power contained in $l+m$=even or odd mode combinations.

\end{abstract}


\section{Introduction}
Isotropy of the universe via the Cosmological principle is one of the foundational
assumptions of modern cosmology. CMB ushered in the precision era in cosmology. 
Consequently it facilitated tests of this otherwise simplifying
assumption of isotropy of cosmos. Multiple studies of CMB data in that direction
indicated instances of isotropy violation.
These were also further validated by the WMAP and Planck CMB collaborations.
See Ref.~\cite{wmap7yranom,wmap9yrmaps,planck13is,planck15is} and also the
references therein.

In this article, we study the anomalous odd power excess in CMB angular power
spectrum, which corresponds to an odd inversion parity preference in the data
\cite{LandMaguiejo05,KimNas10,Gruppuso11,AluriJain12}. 
CMB anisotropies are conventionally expanded in terms of spherical harmonics as
\begin{equation}
\Delta T(\hat{n}) =  \sum_{l=2}^{\infty} \sum_{m=-l}^{+l} a_{lm} \,Y_{lm}(\hat{n})\,,
\end{equation}
where $\Delta T(\hat{n})$ are the fluctuations in CMB temperature anisotropies
around the mean sky temperature ($l=0$) and further removing
the dipole ($l=1$) that is induced due to our relative motion through CMB rest frame.
And $a_{lm}$ are the coefficients of expansion in the spherical harmonic basis, $Y_{lm}(\hat{n})$.
Under inversion, i.e., $\hat{n}\rightarrow-\hat{n}$, the spherical harmonic
coefficients transform as $a_{lm} \rightarrow (-1)^{l}a_{lm}$. Correspondingly
the maps
\begin{equation}
T^\pm(\hat{n}) = \frac{T(\hat{n}) \pm T(-\hat{n})}{2}\,,
\end{equation}
contain only the even/odd modes respectively. Thus, mean power proportional to
$\sum_l C_l$ in some chosen multipole range can be decomposed as $\sum_l C_l = C_l^+ +C_l^- $,
where $C_l = \sum_m |a_{lm}|^2/ (2l+1)$ ($\forall\,\, m=-l$ to $+l$) is the angular power
spectrum of CMB anisotropies, and $C_l^\pm$ is the power in even/odd multipoles respectively.
Excess power in either of these modes will lead to an inversion parity symmetry/asymmetry
preference in the data which is not expected in the standard cosmological model.
This inversion (a)symmetry of  CMB temperature anisotropies was first tested in
Ref.~\cite{LandMaguiejo05} using NASA's WMAP one year data. It was further studied in
Ref.~\cite{KimNas10,Gruppuso11,AluriJain12} using later data releases from WMAP observations.
With new data from a different full-sky mission namely the ESA's PLANCK probe, this anomaly
was still found to persist. Planck collaboration analysed this phenomenon, among others,
to find that its significance varied from 0.002 - 0.004 in the 2013 nominal data
and 0.002 - 0.003 in the 2015 full-mission data depending on the CMB signal extraction method
and galactic masks used~\cite{planck13is,planck15is}.

Alternatively it can be shown that the point parity asymmetry is related to
antipodal correlations of CMB sky~\cite{Nas12}. The two-point correlation function is
defined as
\begin{equation}
C(\theta) = \sum_l \frac{2l+1}{4\pi} C_l P_l(\cos\theta)\,,
\end{equation}
where $C(\theta)=\langle\Delta T(\hat{n})\Delta T(\hat{n}')\rangle$ and
$\hat{n}\cdot\hat{n}'=\cos\theta$.
Taking $\theta=\pi$ corresponding to correlation between anti-podal points
we get,
\begin{eqnarray}
C(\pi) &=& \sum_l \frac{2l+1}{4\pi} C_l P_l(\cos\pi)\\ \nonumber
       &=& \sum_l (-1)^l\frac{2l+1}{4\pi} C_l\\ \nonumber
       &=& \sum_{l_+} \mathcal{C}_l^+ - \sum_{l_-} \mathcal{C}_l^-\,,
\end{eqnarray}
where $\mathcal{C}_l = \frac{2l+1}{4\pi} C_l$ and $\sum_{l_{\pm}} \mathcal{C}_l^\pm$ is the
corresponding power in even/odd only multipoles up to some chosen maximum multipole.
Here $P_l(x)$ are the usual Legendre polynomials.
The odd parity preference i.e., the excess power in odd multipoles compared to even
multipoles in the data is also evident by the negative correlation
seen in the data in the two-point correlation function, $C(\theta)$, for $\theta=\pi$
(see for example Fig.~2 (top left) in Ref.\cite{planck15is}).

Directional dependence of this even-odd multipole power asymmetry was explored
in Ref.~\cite{Nas12}. It was found that the minimum parity asymmetry
i.e., maximum discrepancy in power between even and odd multipoles, was found in the
CMB kinetic dipole direction i.e., ($l,b)=(264^\circ,48^\circ$). A possible relation
between this parity asymmetry in power and other large angular scale CMB anomalies or
features were explored in Ref.~\cite{Nas12,KimNas11,Gruppuso18}.

Further, in the present paper, we also report on any presence of mirror parity (a)symmetry
preference in the data. Presence of any mirror parity preference in CMB data were earlier
pursued in Ref.~\cite{deOliveira04,Bendavid12,Finelli12}, and their possible correlation with
CMB large angle anomalies or features were explored in Ref.~\cite{Nas11,Rassat14}.
Mirror parity was also studied by Planck collaboration in Ref.~\cite{planck13is,planck15is}.
Since mirror (a)symmetry is a statistic defined with respect to the co-ordinate system
used, and we study the CMB maps, which are provided in Galactic
co-ordinates, our results pertain to this choice of the co-ordinate system.

In Ref.~\cite{Aluri17}, collective alignments among even/odd-only multipoles were studied.
Those authors found that the alignments
among even and odd only multipoles span two distinct, non-overlapping regions of the sky
that contain previously known preferred axes corresponding to even/odd parity (a)symmetry or
phenomena such as the axis of maximum mirror parity (a)symmetry, minimum of the even-odd
multipole power asymmetry, hemispherical power asymmetry, and anisotropy axes of $l=2,3$ modes.

The paper is organized as follows. In section~\ref{sec:stat-data} we review the
statistics used in the present work, and also describe the observational data and
complementary simulations used to compute significances. Then in section~\ref{sec:anls},
our results are presented and discussed. Finally we conclude in section~\ref{sec:concl}.


\section{Statistics and Data used}\label{sec:stat-data}
Here we briefly describe the statistics, also the data sets and simulation ensemble
used in the present study. 
The statistics described here will be applied to probe even-odd multipole power (a)symmetry
preference and mirror parity properties of multipoles in the range $l=2-101$ in Planck
full-mission CMB temperature maps i.e., the first 100 cosmological multipoles.

\subsection{Point parity statistic}
We employ the statistics previously defined in Ref.~\cite{KimNas10,AluriJain12},
to probe the even-odd multipole power asymmetry.

In Ref.~\cite{KimNas10}, the power in even multipoles compared to odd multipoles
were analysed by computing the mean power in even and odd multipoles separately up to
a chosen maximum multipole, $l_{max}$, as
\begin{equation}
P^\pm = \sum_{l=2}^{l_{max}} \frac{[1\pm(-1)^l]}{2} \, \mathcal{D}_l\,.
\end{equation}
Here $\mathcal{D}_l=l(l+1)C_l/2\pi$, and $P^\pm$ represents
the mean power in even or odd only multipoles up to a chosen multipole `$l_{max}$'.
Also we take $l_{max}\geq3$. A statistic is then defined by taking their ratio as
\begin{equation}
P(l_{max}) = P^+/P^-\,.
\label{eq:kn-stat}
\end{equation}

Following Ref.~\cite{AluriJain12}, we also use the statistic which is defined as the
average ratio of the power in adjacent even over odd multlipoles
up to that chosen maximum multipole. It is given by
\begin{equation}
Q(l_{odd}) = \frac{2}{l_{odd}-1} \sum_{l=3}^{l_{odd}} \frac{\mathcal{D}_{l-1}}{\mathcal{D}_l}\,,
\label{eq:aj-stat}
\end{equation}
where $l_{odd}$ is the maximum odd multipole up to which the statistic is computed, which will
ensure that there are equal number of even and odd multipoles in the multipole range used to
compute the statistic.
Evidently the summation is taken over only the odd multipoles and
the multipole range involved in this computation is $2 \leq l \leq l_{odd}$.

\subsection{Mirror parity statistic}
Mirror parity corresponds to reflection (a)symmetry with respect to a chosen
plane. Just as even or odd `$l$' modes have point parity (a)symmetry properties,
modes with even or odd `$l+m$' combinations correspond to mirror reflection
(a)symmetry with respect to the equatorial plane of the chosen co-ordinate system
in which the CMB map is represented.

The mirror (a)symmetry was studied first in Ref.~\cite{deOliveira04} using WMAP one year
data. The statistic defined there was adapted and further extended in Ref.~\cite{Finelli12},
to find significant odd mirror parity preference in WMAP's seven year data. Independently,
this (a)symmetry was also probed in Ref.~\cite{Bendavid12} to again find a significant odd
mirror parity preference. The same was assessed using Planck data and an odd mirror parity
was found to be present in it as well \cite{planck13is,planck15is}.
 
The mirror parity statistics used in the present work are summarized below :
\begin{eqnarray}	
\label{eq:mirr-stat}
 E_l&=&\frac{\langle|a_{lm}^2|\rangle_{l+m=even}}{C_l^{th}}\\ \nonumber
 O_l&=&\frac{\langle|a_{lm}^2|\rangle_{l+m=odd}}{C_l^{th}}\\ \nonumber
 R_l&=&\frac{\langle|a_{lm}^2|\rangle_{l+m=even}}{\langle|a_{lm}^2|\rangle_{l+m=odd}} \\ \nonumber
 D_l&=&\frac{\langle|a_{lm}^2|\rangle_{l+m=even}-\langle|a_{lm}^2|\rangle_{l+m=odd}}{C_l^{th}}\,,
\end{eqnarray}
where $\langle|a_{lm}^2|\rangle_{l+m=even/odd}$ denotes the mean power in even or odd
spherical harmonic coefficients for the combination $l+m$=even/odd respectively.
The $E_l$ and $O_l$ statistics separately probe excess/deficit of power in even/odd
$l+m$ modes of a particular `$l$' in the data with respect to expected power
in standard concordance model. $R_l$
and $D_l$ probe the ratio of and difference between the power in even mirror parity modes
compared to odd mirror parity modes, respectively. All these statistics
reveal in various ways how the power is distributed between even and odd mirror parity
modes. Here $C_l^{th}$ is the fiducial power spectrum based on best fit $\Lambda$CDM
cosmological parameters from Planck 2015 data.

One can also define these mirror statistics with respect to the power seen in
a particular realization in a multipole i.e., using $C_l^{data}$ instead of $C_l^{th}$.
In that case $E_l = {\langle|a_{lm}^2|\rangle_{l+m=even}}/{C_l^{data}}$ will be the
only statistic that can be defined independently. The other statistics can be
defined in terms of $E_l$ as $O_l=1-E_l$ and $D_l = 2E_l-1$. The $R_l$ statistic
will remain unaltered, and hence no new information will be furnished by it when we replace
$C_l^{th}$ by $C_l^{data}$. In this work, however, we will not pursue this.

\subsection{Data}

\begin{figure}[t]
\centering
\includegraphics[width=0.55\textwidth]{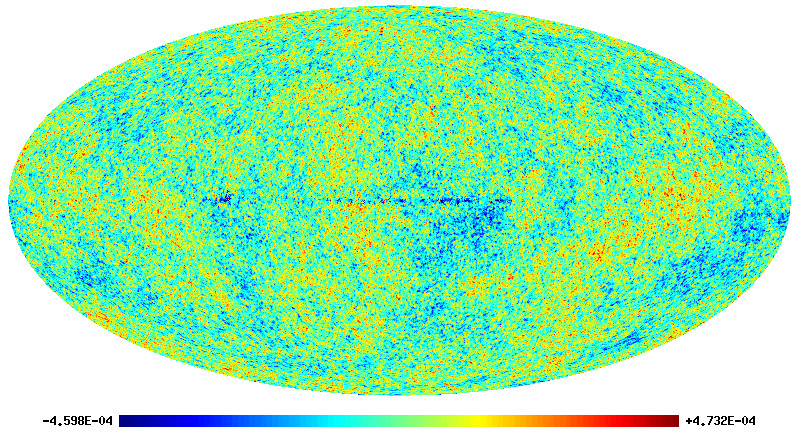}
~
\includegraphics[width=0.48\textwidth]{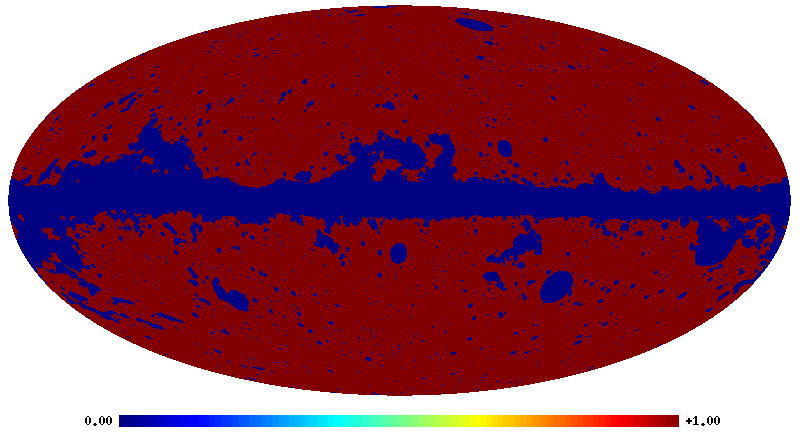}
~
\includegraphics[width=0.48\textwidth]{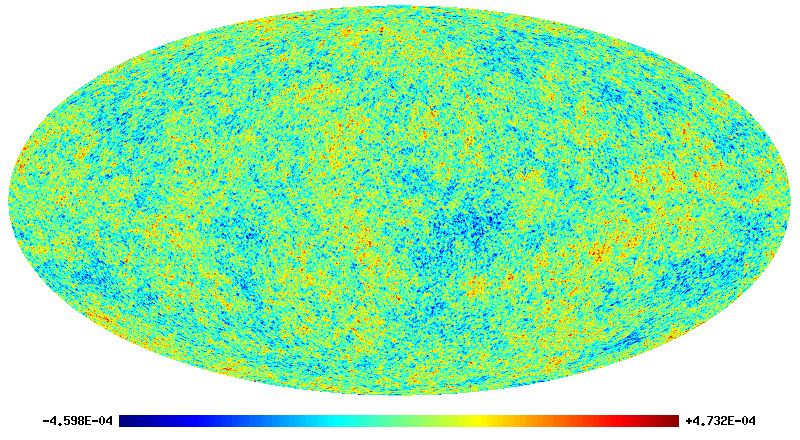}
~
\includegraphics[width=0.48\textwidth]{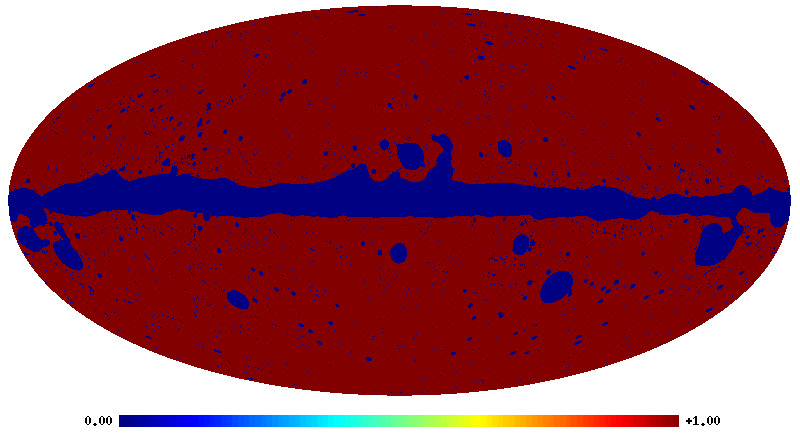}
~
\includegraphics[width=0.48\textwidth]{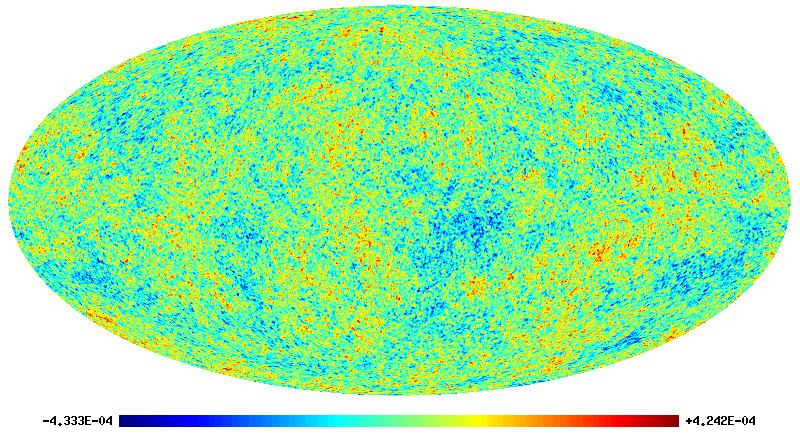}
\caption{ \emph{Top :} Cleaned SMICA map provided by Planck team, 
\emph {Middle left :} UT78 common temperature mask, 
\emph{Middle right :} Inpainted SMICA map masked with UT78 mask, 
\emph{Bottom left :} SMICA component separation specific mask, 
\emph{Bottom right :}  Inpainted SMICA map masked with SMICA mask. 
All maps are in $N_{side}=2048$ }
\label{fig:cmb-map-masks}
\end{figure}

For the current study, we use the CMB data from Planck 2015 public release. Specifically
we use the SMICA 2015 CMB map. The SMICA map as provided by the Planck 2015 Collaboration
is shown in \emph{top} panel of Fig.~[\ref{fig:cmb-map-masks}].
It is available at an $N_{side}=2048$, where $N_{side}$ is the
\texttt{HEALPix}\footnote{\url{https://healpix.jpl.nasa.gov/}} map resolution
parameter. Although one employs a cleaning procedure to recover the pristine cosmic signal
of our interest i.e., CMB from the raw satellite data, there is still non-negligible
contamination present in the recovered CMB map particularly in the galactic plane
owing to strong foreground signal. Thus a portion of the recovered CMB sky
is omitted i.e., masked, before undertaking any cosmological analysis, which
can potentially bias our inferences.

In this work we use two masks viz., the
UT78 common analysis mask and the SMICA mask. The UT78 mask is the most conservative mask
to omit foreground residuals. Planck collaboration employed four different cleaning
methods to recover cosmic CMB signal~\cite{planck15cmb}. UT78 mask can be used with any of the
four CMB maps thus derived using those four cleaning procedures.
As the name suggests, it has a usable sky fraction (usually referred to as $f_{sky}$)
of approximately $78\%$, and is the union of all the galactic masks specific to
each component separation method.
In addition, we also use the SMICA mask - specific to SMICA CMB recovery
procedure - which defines the region of its efficiency in reducing foregrounds.
Although it is sufficient to use SMICA mask,
we use these two masks to compare the effect of various levels of galactic cut
on our statistics, employed to study the power and mirror parity (a)symmetry
in the present work. The UT78 and SMICA masks are shown in \emph{middle left}
and \emph{bottom left} panels of Fig.~[\ref{fig:cmb-map-masks}].

To study the mirror parity (a)symmetry, we need the individual spherical harmonic
coefficients ($a_{lm}$'s) of full-sky CMB map to compute power in any set of $l+m$=even
or odd modes separately. Thus we apply an \emph{inpainting} method, which can be thought
of as a $2D$ interpolation on the sphere, on the data map to recover full-sky CMB map
from a partial sky, left after masking for omitting foreground residuals.
The $a_{lm}$'s are thus extracted from the inpainted/pseudo full-sky SMICA CMB map,
and tested for even/odd/no mirror parity preference in the data.
Since inpainting takes a lot of time ( $\sim$ 180 minutes with
a machine using Intel CORE M Processor, 4 CPUs (2 Threads per core) and 4 GB of RAM)
to fill an $N_{side}=2048$ map, and further
since we are also only interested in studying the low multipole
regime $l = [2, 101]$, we only apply inpainting on the data maps.
These masked, inpainted maps are used to assess point parity preferences
in the data  i.e., we use $C_l$ obtained from these inpainted (pseudo) full-sky CMB maps.
The inpainted SMICA CMB map as masked and inpainted using the two masks - UT78 and SMICA mask -
are shown in \emph{middle right} and \emph{bottom right} panels of Fig.~[\ref{fig:cmb-map-masks}],
respectively. We use the freely available
\texttt{iSAP}\footnote{\url{http://www.cosmostat.org/software/isap}. The default options were used.}
software package to inpaint masked SMCIA CMB map.

The significance of our parity statistics is estimated by comparing the data estimated
values (both cleaned map as provided and the two inpainted SMICA maps) with 1000 mock CMB maps
added with appropriate noise.
The random CMB realizations were generated as follows :
\begin{itemize}
\item
We first generate the best-fitting theoretical angular power spectrum ($C_l$)
using Planck 2015 best-fit cosmological parameters~\cite{planck15cosmopar}
as input to \texttt{CAMB} software\footnote{\url{https://camb.info/}}.
\item
Using these theoretical $C_l$, we generated 1000 random CMB realizations with
a beam resolution of a Gaussian beam of $FWHM=5'$ (arcmin).
\item
Each of these CMB maps are then added with 1000 SMICA-like noise realization
obtained from Planck public release 2\footnote{\url{http://crd.lbl.gov/cmb-data}}.
\end{itemize}

The significance of parity statistics defined in Eq.~[\ref{eq:kn-stat}], [\ref{eq:aj-stat}] and
[\ref{eq:mirr-stat}] are estimated by comparing the data statistics with 1000 simulations generated
as mentioned above. We compare the cleaned SMICA map as provided
by the Planck team, and also after inpainting with the two masks.
Thus we have three sets of values for each statistic from data, to compare with
simulations.


\section{Results}\label{sec:anls}
\subsection{Power parity and it's statistical significance}

\begin{figure}[!t]
\centering
\includegraphics[width=0.9\textwidth]{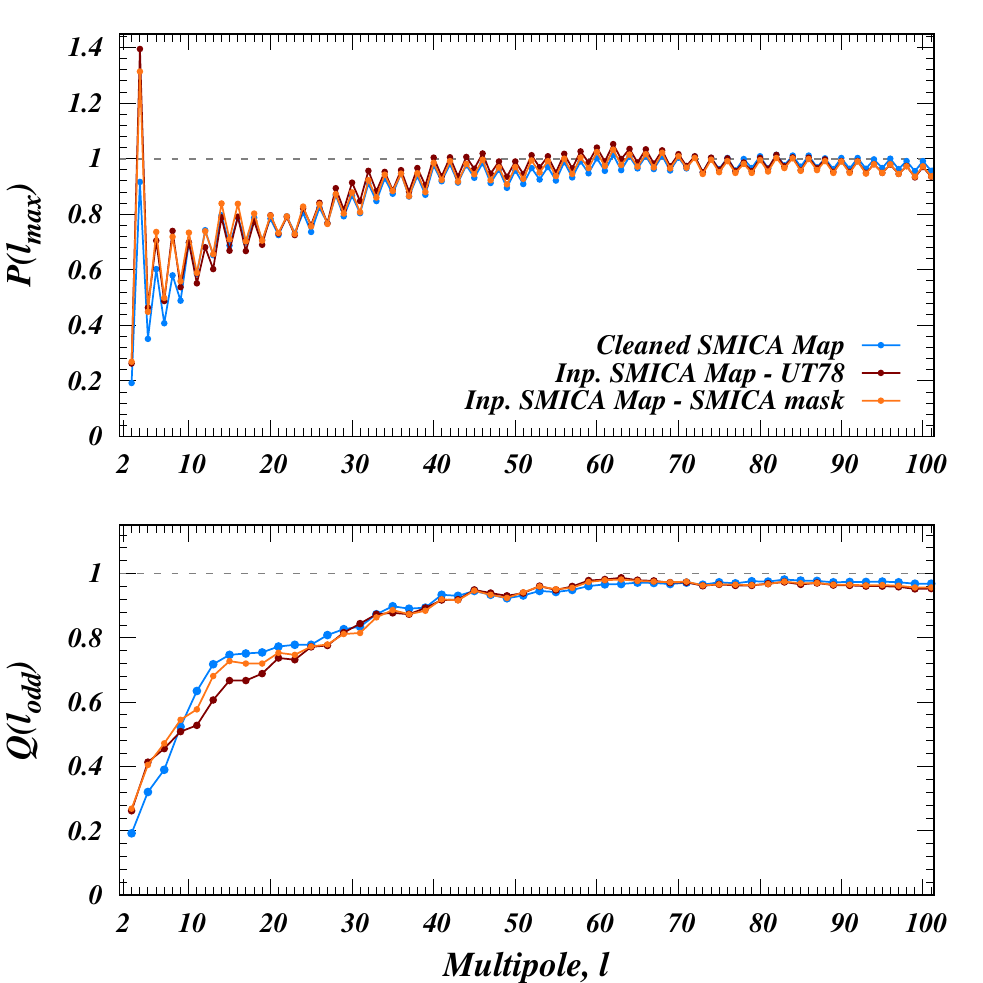}
\caption{The even-odd multipole power asymmetry in the SMICA temperature map in 
the multipole range $l=[2,101]$. The asymmetry is computed using both the $P(l_{max})$
(\emph{top panel}) and $Q(l_{odd})$ (\emph{bottom panel}) statistics. Each panel has
three data curved corresponding to the cleaned SMCIA map as provided by Planck collaboration
(blue), inpanited SMICA map after masking with UT78 mask (maroon), and inpainted SMICA CMB
map after masking with SMICA component separation specific mask (orange).} 

\label{fig:kn-aj-stat-data}
\end{figure}

As mentioned before, we have three sets of data values for each statistic as obtained from
the CMB map as is provided, inpainted SMICA map using the UT78 common mask, and finally
the inpainted SMICA map obtained using the SMICA mask, both generated at $N_{side}=2048$.

The parity statistic values from data, using $P(l_{max})$ and $Q(l_{odd})$ as defined
in Eq.~[\ref{eq:kn-stat}] and [\ref{eq:aj-stat}] are shown in \emph{top} and \emph{bottom}
panels of Fig.~[\ref{fig:kn-aj-stat-data}], respectively.
Note that the $C_l$ obtained from data are deconvolved with appropriate beam
and pixel window functions in harmonic space, before computing the statistics.
One can readily notice that, there is excess power
in odd multipoles compared to even multipoles as both the statistics are $<1$ in the range
of multipoles we studied. This is also
consistent with previous findings in Ref.~\cite{LandMaguiejo05,KimNas10,Gruppuso11,AluriJain12}.
The three set of statistic values for each kind of data map generated as mentioned earlier,
are displayed in each panel in different colours.

\begin{figure}[!t]
\centering
\includegraphics[width=0.9\textwidth]{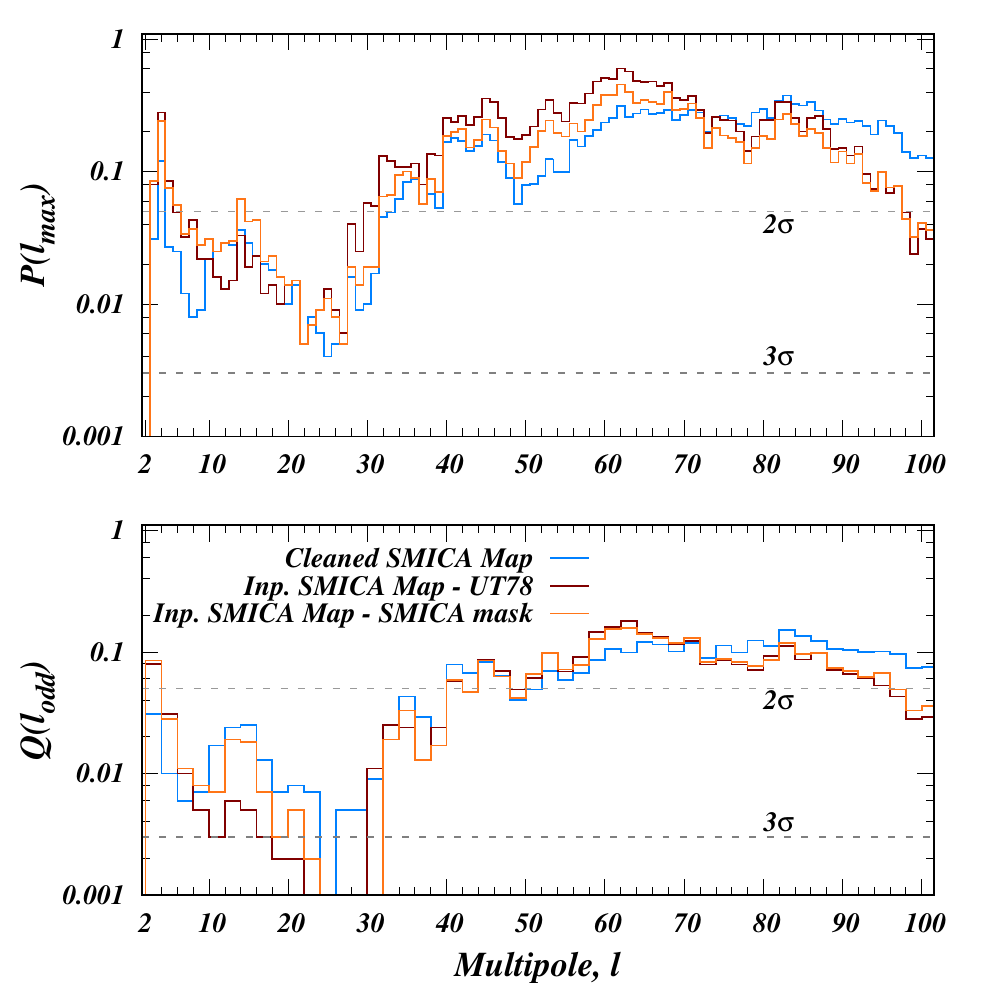}
\caption{Probability estimates of parity asymmetry in data as quantified by the statistics
$P(l_{max})$ and $Q(l_{odd})$ used in our study. This plot is organized and color coded
in the same way as Fig.~[\ref{fig:kn-aj-stat-data}]. The data statistic values were compared
with 1000 simulations to compute these $p$-values.}
\label{fig:kn-aj-stat-sign}
\end{figure}

In order to compare the data statistic values, we compute the same quantities
from our simulation ensemble of 1000 SMICA-like maps,
in the same way as the data statistic values are obtained.
The corresponding significances or $p$-values are shown in Fig.~[\ref{fig:kn-aj-stat-sign}].
The \emph{top} panel of Fig.~[\ref{fig:kn-aj-stat-sign}]
corresponds to significances for the statistic $P(l_{max})$,
and \emph{bottom} panel of the same figure corresponds to significances for the statistic
$Q(l_{odd})$. Again the significances corresponding to the three types of data maps generated
for this study, are shown in the same way as Fig.~[\ref{fig:kn-aj-stat-data}].

These significance plots reveal that the $p$-values found here are similar to those reported
previously in Ref.~\cite{AluriJain12}. The pattern and level of significance of excess power in odd
multipoles compared to even multipoles for any of the statistics used, are almost the same as before.
The significances of both $P(l_{max})$ and $Q(l_{odd})$ statistics reach a maximum at around
$l_{max}$ or $l_{odd} \sim 29$. However the $Q(l_{odd})$ statistic shows an almost $\sim 3\sigma$
significance compared to $P(l_{max})$ statistic which reveals only a $2\sigma$ discrepancy in the
data at this maximum multipole $\sim 29$. Further the $Q_{l_{odd}}$ statistic continues to have
significances with probability $p\lesssim 2\sigma$ up to
$l_{odd}\lesssim 49$, and $p \lesssim 0.9$ beyond that. These trends are again similar to our
findings in Ref.~\cite{AluriJain12}.

We also note that the three statistic values corresponding to the three types of data
created for our study from PR2, have a similar level of significances. Furthermore, we
used the more conservative UT78 mask as well as the SMICA CMB map specific component
separation mask. This gives confidence to our study that the significances found here
cannot be attributed to foreground residuals. The cumulative statistic values at low
multipoles are low, that can be arising due to large cosmic variance resulting in
wider excursion of statistics' values.

The $p$-values shown in Fig.~[\ref{fig:kn-aj-stat-sign}] that lie below $0.001$ are
those for which none of
the corresponding statistic values computed from simulations are smaller than the
data statistic value. So their $p$-value should be treated as `$<1$ in $1000$' and not `$0$'.
Note that since both the statistics are symmetry based, they are as blind or a priori as one
chooses to refer them to be.

\subsection{Mirror parity and its statistical significance}
The mirror parity (a)symmetry statistics as defined in Eq.~[\ref{eq:mirr-stat}] are coordinate
dependent. In our analysis, we use the $a_{lm}$'s from Planck CMB temperature maps in the same
co-ordinate system as provided. Thus our probes of preference or no preference for mirror
(a)symmetry from the significances eventually found, are specific to that coordinate system.

Under the assumption of isotropy of the standard cosmological model, the power distribution between
even and odd $l+m$ modes should be same i.e., their ratio should be $\sim 1$ or difference should
be $\sim 0$. This assumption of isotropy further implies that the distribution
of power between even and odd mirror parity modes should be statistically the same in any co-ordinate
system we choose to represent the CMB maps. In the present work, we confine ourselves to the analysis
of maps as provided in the galactic coordinate system. From the assumption of statistical isotropy,
$E_l$, $O_l$ and $R_l$ are expected to vary about $\sim 1$, and $D_l$ about $\sim 0$.

Thus large deviations about $\sim 1$ (or $\sim 0$ depending on the statistic) would then suggest
a deviation from isotropy. How much ever the deviation, large or small,
the statistic values may \emph{seem} to be from `$1$' (or $\sim 0$), their significances have to
be computed using appropriately generated corresponding simulations.

\begin{figure}[!t]
\centering
\includegraphics[width=0.9\textwidth]{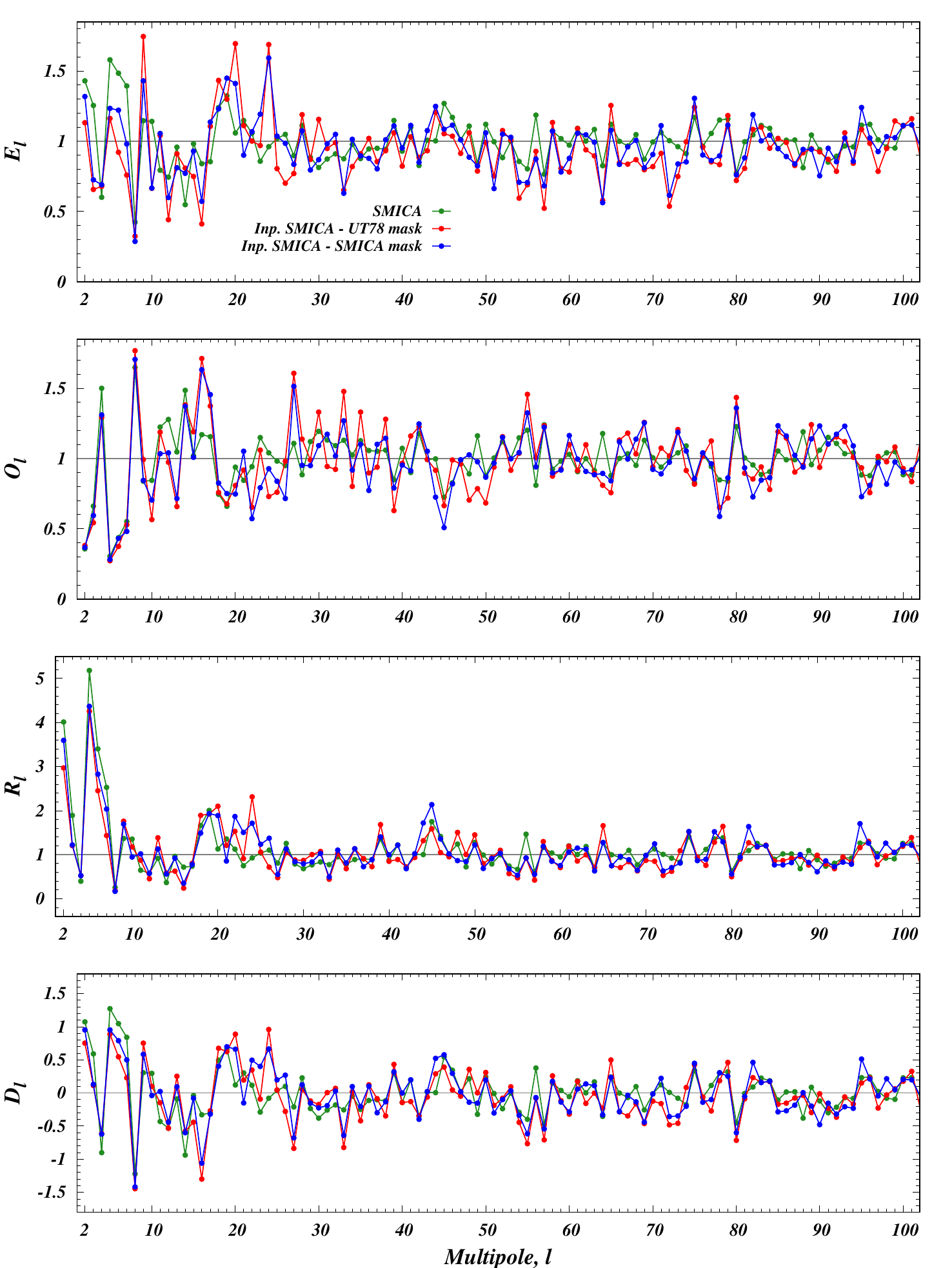}
\caption{Shown here are the values of mirror pairty statistics $E_l$, $O_l$, $R_l$ and $D_l$
computed from data from top to bottom respectively. Each panel has the statistic value
computed for SMICA map as is provided (green), inpainted SMICA map after applying UT78 mask (red),
and inpainted SMICA map after applying SMICA mask (blue). The horizontal solid gray line in each
panel is the theoretically expected value for the statistic.}
\label{fig:mirror-stat-data}
\end{figure}

The mirror statistics' values as found for the SMCIA 2015 CMB map are shown in
Fig.~[\ref{fig:mirror-stat-data}]. All the three types of data viz., the cleaned
SMICA map as provided, SMICA map inpainted using both the UT78 mask and SMICA component
specific mask are shown in \emph{green}, \emph{red} and \emph{blue} colours
respectively. One can see that any of the statistics as defined in Eq.~[\ref{eq:mirr-stat}]
has values fluctuating around the expected theoretical value that is shown by a horizontal
\emph{gray} solid line.

\begin{figure}[!t]
\centering
\includegraphics[width=0.9\textwidth]{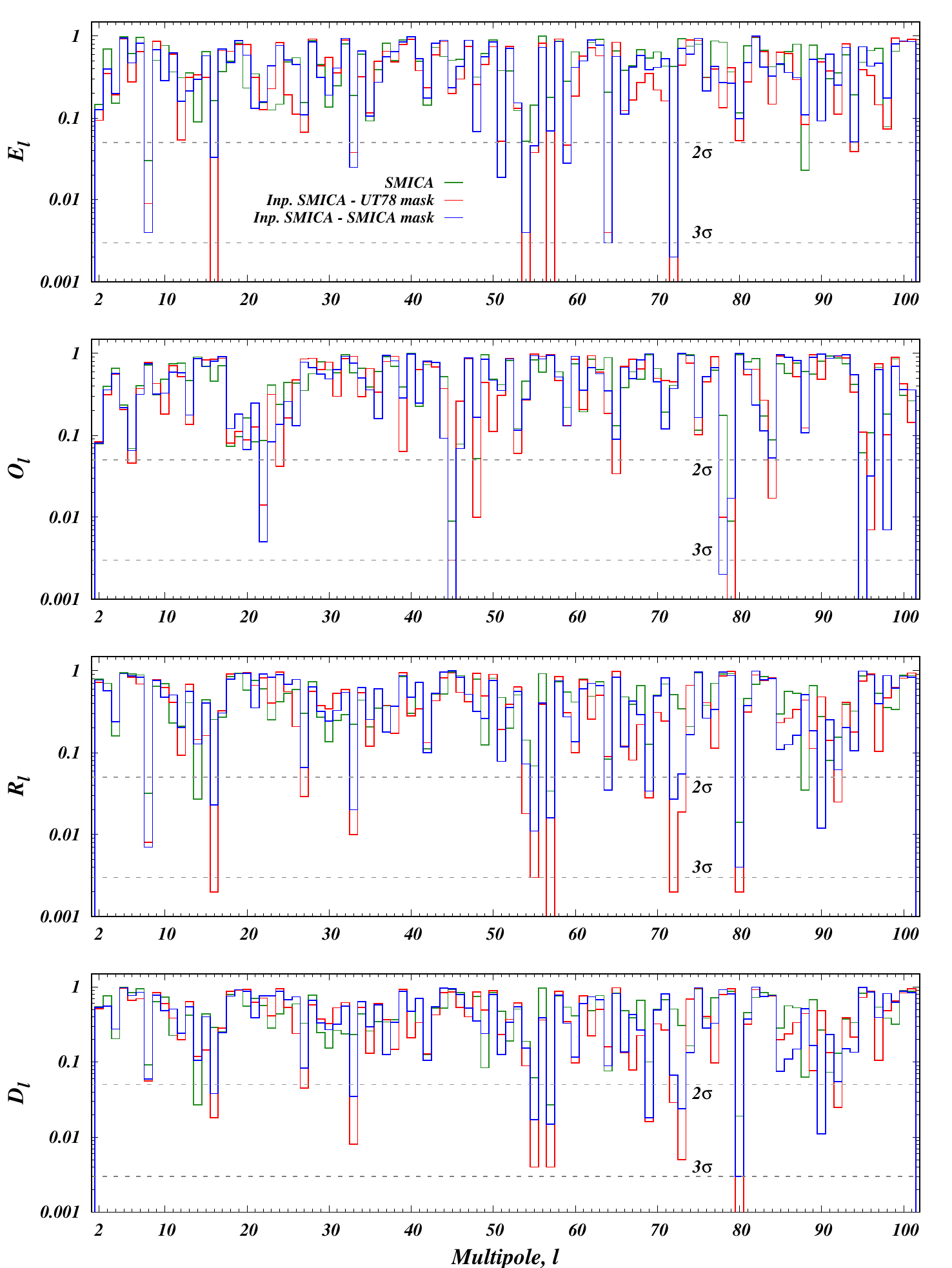}
\caption{Probability estimates of mirror parity statistics $E_l$, $O_l$, $R_l$
and $D_l$ are shown here for SMICA map (green), inpainted SMICA map after applying UT78 mask (red),
and inpainted SMICA map after applying SMICA procedure specific confidence mask (blue)
in each panel.}
\label{fig:mirror-stat-sign}
\end{figure}

The significance of various Mirror parity statistics are calculated by comparing with
1000 simulated SMICA-like CMB maps.
The observed significances are shown in Fig.~[\ref{fig:mirror-stat-sign}] for each of the
statistics $E_l$, $O_l$, $R_l$ and $D_l$ in first, second, third and fourth panels from
the top respectively. Except for few multipoles, no significant mirror parity preference was
found in the data with any of the statistics we used, in the range $l=[2,101]$ that we studied.
We recall that these maps are
in galactic coordinates, and the mirror parity statistics we used (as defined in
Eq.~[\ref{eq:mirr-stat}]) are coordinate dependent. So these results are specific to
that coordinate system.

The fundamental assumption of the current standard model of cosmology, the
Cosmological Principle, specifically isotropy, implies that there should not be any
preference for mirror (a)symmetry independent of the coordinate system in which a
CMB map is represented. However, in a given realization
there will always be an axis along which more power is distributed in even or odd
`$l+m$' modes. This observed level of discrepancy has to be compared with
simulations. Indeed it was found that there is an anomalous odd mirror parity
preference in the data with respect to the direction $(l,b)\approx (264^\circ,-17^\circ)$
\cite{Finelli12,Bendavid12} in the WMAP data at a significance of $>3\sigma$.
It was also found to be anomalous in the Planck data as well, in almost the
same direction with a significance of $\sim 99\%$. We explore this anomalous
nature of odd mirror parity preference in the data and its direction
further lateron.


\section{Conclusions}\label{sec:concl}
In this paper, we revisited our previous work \cite{AluriJain12} on the anomalous
power excess in odd over even multipoles in WMAP data. Using two different statistics
as before, we studied the CMB signal as recovered from Planck full-mission data, specifically
using the SMICA CMB map. We also analysed the data for any mirror
parity (a)symmetry preference in the CMB sky.

Our analysis was performed on the SMICA map as is provided in Galactic coordinates.
Further, to understand the effect of foreground residuals vis-a-vis galactic cuts
on our results,
we also made use of pseudo full-sky SMICA maps obtained by inpainting using
\texttt{iSAP} software, after masking with UT78 common mask and the component
separation specific SMICA temperature mask at $N_{side}=2048$ map.
Mock CMB realizations numbering 1000 were generated with a resolution of Gaussian
beam of $FWHM=5'$ (arcmin). Then, these maps were added with FFP8.1 SMICA noise maps
that are made publicly available through Planck PR2.

The objective of our present work was to probe point and mirror parity (a)symmetry
preferences or consistency with standard model expectations in the latest
CMB data from Planck. Appropriate statistics were used to probe any presence
of these (a)symmetries. The data is expected to show no preference for odd
over even point or mirror parity and vice versa.

An odd point parity preference i.e., more power
in odd multipoles compared to even multipoles, is found to persist even in
the Planck full-mission data. Both the statistics, defined in Eq.~[\ref{eq:kn-stat}] and
[\ref{eq:aj-stat}], become maximally anomalous for the multipole range
$l\approx[2,29]$. The statistic values and their significances show the same
pattern and level as found previously in Ref.~\cite{AluriJain12} using WMAP data.

We then studied the full-mission SMICA CMB temperature map for the
presence of any mirror parity preferences in the
data using the statistics defined in Eq.~[\ref{eq:mirr-stat}]. Since these
are coordinate dependent quantities, our results pertain to the CMB data
represented in Galactic coordinates, as provided by Planck collaboration.
We find no significant even or odd mirror parity preferences in the data
at any multipole `$l$' except for few random modes. Thus with respect to
the galactic plane, the SMICA CMB map from Planck PR2 doesn't display
any significant mirror parity (a)symmetry. In the future we plan to pursue
the frame dependence of mirror parity in more detail.

\section*{Acknowledgements}
Some of the results in the current work were derived using the publicly
available \texttt{HEALPix} package~\cite{Healpix}. We also acknowledge
the use of \texttt{CAMB}\footnote{\url{https://camb.info/}}, a
freely available Boltzmann solver for CMB anisotropies. Part of
the results presented here are based on observations obtained with
Planck\footnote{\url{https://www.cosmos.esa.int/web/planck}}, an ESA
science mission with instruments and contributions directly
funded by ESA Member States, NASA, and Canada.
This work also made use of \texttt{iSAP} software~\cite{isap}.
SP acknowledges DST-INPIRE for financial support under the research
grant DST/INSPIRE/03/2014/004358, and PKS acknowledges Utkal University
for financial support under university seed research grant.

\end{document}